\documentclass[12pt]{article}

\catcode`\@=11
\@addtoreset{equation}{section}

\global\arraycolsep=2pt

\usepackage{mathrsfs,amsbsy,amssymb,latexsym,amsfonts,amsmath,cite}
\usepackage{graphicx,color}

\allowdisplaybreaks

\newcommand{\al}{\alpha}
\newcommand{\be}{\beta}
\newcommand{\ga}{\gamma}

\newcommand{\ep}{\epsilon}

\newcommand{\ze}{\zeta}

\newcommand{\Tr}{{\rm Tr}}

\newcommand{\alg}[1]{\mathfrak{#1}}
\newcommand{\el}{\nonumber}

\begin{document} 
% \maketitle
% \flushbottom

\begin{flushright}
\parbox{4cm}
{KUNS-2488 \\ 
ITP-UU-14/12 \\ 
SPIN-14/12}
\end{flushright}

\vspace*{1.5cm}

\begin{center}
{\Large\bf Lunin-Maldacena backgrounds \\ 
\qquad from the classical Yang-Baxter equation} 
\vspace*{0.3cm}\\ 
- {\it\large Towards the gravity/CYBE correspondence} - 
\vspace*{1.5cm}\\
{\large Takuya Matsumoto$^{\dagger}$\footnote{E-mail:~t.matsumoto@uu.nl} 
and Kentaroh Yoshida$^{\ast}$\footnote{E-mail:~kyoshida@gauge.scphys.kyoto-u.ac.jp}} 
\end{center}
\vspace*{0.25cm}
\begin{center}
$^{\dagger}${\it Institute for Theoretical Physics and Spinoza Institute, 
Utrecht University, \\ 
Leuvenlaan 4, 3854 CE Utrecht, The Netherlands.} 
\vspace*{0.25cm}\\ 
$^{\ast}${\it Department of Physics, Kyoto University, \\ 
Kyoto 606-8502, Japan.} 
\end{center}
\vspace{1cm}

\begin{abstract}
We consider $\gamma$-deformations of the AdS$_5\times$S$^5$ superstring 
as Yang-Baxter sigma models with classical $r$-matrices 
satisfying the classical Yang-Baxter equation (CYBE). 
An essential point is that the classical $r$-matrices are composed of Cartan generators only 
and then generate abelian twists. 
We present examples of the $r$-matrices that lead to real $\gamma$-deformations of 
the AdS$_5\times$S$^5$ superstring. Finally we discuss a possible classification of 
integrable deformations and the corresponding gravity solution in terms of solutions of CYBE. 
This classification may be called the gravity/CYBE correspondence. 
\end{abstract}

\setcounter{footnote}{0}
\setcounter{page}{0}
\thispagestyle{empty}

\newpage

\section{Introduction}

The AdS/CFT correspondence \cite{M,GKP,W} has been well investigated  
and there would be no doubt for its validity at least in the planar limit. 
However, it is still important to consider the fundamental structure of the duality 
in order to make our understanding much deeper and look for a clue of new physics. 
The discovery of the integrable structure behind it \cite{review} 
would play an important role along this direction. 
The integrability provides a guiding principle to extend the AdS/CFT correspondence 
by elaborating integrable deformations of it.  

\medskip 

Our concern here is the integrable structure of type IIB superstring on AdS$_5\times$S$^5$\,. 
The Green-Schwarz string action is constructed from the following supercoset \cite{MT}: 
\[
PSU(2,2|4)/\left[SO(1,4) \times SO(5)\right]\,. 
\] 
The $\mathbb{Z}_4$-grading of this coset ensures the classical integrability \cite{BPR}. 
For similar argument based on another coset representation \cite{RS}, see \cite{Hatsuda}. 
Possible supercosets, which lead to classically integrable and consistent string theories, are classified 
in \cite{Zarembo-symmetric,Wulf}.  

\medskip 

A recent interest is to consider $q$-deformations of the AdS$_5\times$S$^5$ superstring. 
There are two kinds of $q$-deformations, 1) standard $q$-deformations \cite{Drinfeld1,Drinfeld2,Jimbo} 
and 2) non-standard $q$-deformations (also called Jordanian deformations) \cite{Jordanian,KLM}. 
Both of them are based on the Yang-Baxter sigma model approach proposed by Klimcik 
\cite{Klimcik}\footnote{For quantum aspects of Yang-Baxter sigma models, see \cite{Squellari}. }, 
where linear R-operators constructed from classical $r$-matrices play the central role 
in constructing deformed classical actions.  
As a characteristic property, the former is based on the modified Yang-Baxter equation (mCYBE) and 
the latter is on the classical Yang-Baxter equation (CYBE). 

\medskip 

For standard $q$-deformations of sigma models, many works have been done so far. 
Although deformed target spaces are not represented by symmetric cosets\footnote{
For some examples of non-symetric cosets, see \cite{SYY}.} 
and there is no general prescription to argue the integrability,  
many techniques have been developed and various aspects 
have been revealed. Especially for squashed S$^3$\,, 
the Lax pair was presented in \cite{Cherednik} 
and the classical integrable structure has been elaborated in the subsequent works 
\cite{FR,BFP,Mohammedi,KY,KYhybrid,KMY-QAA,KMY-monodromy,ORU,KOY,BR}. 
As a possible way toward higher-dimensional cases, the Yang-Baxter sigma model approach was proposed by 
Klimcik \cite{Klimcik}. Though it was originally argued for principal chiral models, 
Delduc, Magro and Vicedo succeeded to generalize it to symmetric coset cases \cite{DMV}, 
where the standard $q$-deformed algebra is presented as a generalization of \cite{KYhybrid}. 
Then they have constructed a standard $q$-deformed AdS$_5\times$S$^5$ superstring action 
with a linear R-operator satisfying mCYBE \cite{DMV2}. The coordinate system was introduced 
and the metric in the string frame and NS-NS two-form have been determined so far \cite{ABF}. 
However the full solution has not been obtained yet in type IIB supergravity. 
For further discussion with specific values of the deformation parameter, 
see \cite{HRT}. A related mirror TBA is also discussed in \cite{AdLvT}. 
It would be an important task to compare the results with the deformed S-matrices 
\cite{BK,HHM,dLRT,Arutyunov}. 

\medskip 

For non-standard $q$-deformations,  
deformed AdS$_5\times$S$^5$ superstring actions have been constructed  
with linear R-operators satisfying CYBE \cite{KMY-Jordanian-typeIIB}. 
A remarkable point is that partial deformations are possible 
in comparison to the standard $q$-deformation. 
For a simple example of the classical $r$-matrices deforming only AdS$_5$\,, 
the metric and NS-NS two-form are obtained by a coset construction with an appropriate coordinate system. 
Then the complete type IIB gravitational solution has been found \cite{SUGRA-KMY}. 
In particular, the solution is real and there is no curvature singularity, while the tidal force 
diverges at the boundary except for a specific surface. It also contains the three-dimensional Schr\"odinger spacetime 
as a subspace and, for the subsector analysis, one can use the results obtained in a series of works 
\cite{KY-Sch,Jordanian-KMY,Kame}. All of the results are consistent to the recent analysis \cite{GS}.

\medskip 

In this note, we consider $\gamma$-deformations of the AdS$_5\times$S$^5$ superstring 
as Yang-Baxter sigma models with classical $r$-matrices satisfying CYBE. 
An essential point is that the classical $r$-matrices are composed of Cartan generators only 
and do not satisfy the nilpotency condition in comparison to Jordanian deformations 
considered in \cite{KMY-Jordanian-typeIIB}. These generate abelian twists which are particular examples 
of the Drinfeld-Reshetikhin twists \cite{Drinfeld1,Drinfeld2,R}.  
We present examples of the $r$-matrices that lead to real $\gamma$-deformations of 
the AdS$_5\times$S$^5$ superstring. Finally we discuss a possible classification of 
integrable deformations and the corresponding gravity solution in terms of solutions of CYBE. 
This classification may be called the gravity/CYBE correspondence. 

\medskip 

This note is organized as follows. 
In section 2 we introduce skew-symmetric classical R-operators composed of Cartan generators only. 
These are solutions of CYBE, but do not satisfy the nilpotency condition. 
With the R-operators, we present classically integrable and $\kappa$-invariant string actions. 
Section 3 presents simple examples.  
We show a relation between a classical $r$-matrix and a TsT transformation 
which leads to a real $\gamma$-deformed AdS$_5\times$S$^5$\,.  
It is straightforward to generalize the classical $r$-matrix for three-parameter $\gamma$-deformations.  
As a result, the Lunin-Maldacena background is contained as a particular case.  
Section 4 is devoted to conclusion and discussion. 
Appendix A explains our notation and convention. In appendix B the metric of the three-parameter 
$\gamma$-deformed AdS$_5\times$S$^5$ is rewritten for our convenience. 
Appendix C describes in detail the derivation of the metric and the NS-NS two-form from the Yang-Baxter 
sigma model approach.

\section{Abelian twists of the AdS$_5\times$S$^5$ superstring} 

In this section, after reviewing the formulation of the Yang-Baxter sigma models for  
the AdS$_5\times$S$^5$ superstring action with CYBE \cite{KMY-Jordanian-typeIIB}\,, 
we consider a particular class of $r$-matrices composed of Cartan generators only. 
These generate abelian twists, which are examples of the Drinfeld-Reshetikhin twists \cite{Drinfeld1,Drinfeld2,R}.

\subsection{Deformed AdS$_5\times$S$^5$ string actions with CYBE \label{2.1}}

We are concerned here with the deformed Green-Schwarz string action \cite{KMY-Jordanian-typeIIB}, 
\begin{eqnarray}
S=-\frac{1}{4}(\ga^{\al\be}-\ep^{\al\be})\int^\infty_{-\infty}\!\!\!d\tau\int^{2\pi}_0\!\!\!d\sigma~
{\rm Str}\Bigl(A_\al d\circ\frac{1}{1-\eta R_g\circ d}A_\be\Bigr)\,,  
\label{action}
\end{eqnarray}
where the left-invariant one-form $A_\alpha$ is defined as 
\begin{eqnarray}
A_\alpha\equiv g^{-1}\partial_\alpha g\,, \qquad 
g\in SU(2,2|4)\,. 
\end{eqnarray}
Here $\ga^{\al\be}$ and $\ep^{\al\be}$ are the flat metric and the 
anti-symmetric tensor on the string world-sheet.  
The operator $R_g$ is defined as 
\begin{align}
R_g(X)\equiv g^{-1}R(gXg^{-1})g\,,  
\end{align} 
where a linear operator $R$ satisfies CYBE rather than mCYBE \cite{DMV2}\,. 
The $R$ operator is related to the tensorial representation of classical $r$-matrix through 
\begin{align}
R(X)=\Tr_2[r(1\otimes X)]=\sum_i a_i\Tr(b_iX)
\quad \text{with}\quad r=\sum_i a_i\otimes b_i\,. 
\end{align}
The operator $d$ is given by the following,  
\begin{align}
d=P_1+2P_2-P_3\,,  
\label{d}
\end{align}
where $P_i$ ($i=0,1,2,3$) are the projections to the $\mathbb{Z}_4$-graded 
components of $\alg{su}(2,2|4)$\,.  
$P_0\,,P_2$ and $P_1\,,P_3$ are the projectors to the bosonic and 
fermionic generators, respectively. 
In particular, $P_0(\alg{su}(2,2|4))$ is nothing but $\alg{so}(1,4)\oplus\alg{so}(5)$\,.

\medskip 

For the action \eqref{action} with a Jordanian R-operator, the Lax pair has been constructed \cite{KMY-Jordanian-typeIIB} 
and the classical integrability is ensured in this sense. 
The $\kappa$-invariance has been proven as well \cite{KMY-Jordanian-typeIIB}\,. 
Here it is worth noting that the nilpotency condition is not necessary 
for the $\kappa$-invariance and the classical integrability, though it is a sufficient condition 
to ensure the existence of $1/(1-\eta R_g \circ d)$\,. 
This will be a key observation for our later discussion.

\subsection{Classical R-operators for abelian twists} 

In the previous work \cite{KMY-Jordanian-typeIIB}, we have studied 
classical $r$-matrices of Jordanian type, which satisfy the following properties: 
1) solutions of the classical Yang-Baxter equation (CYBE), 2) the skew-symmetricity, 3) the nilpotency. 
The nilpotency condition is a characteristic property of Jordanian type.  
A simple example to deform only the AdS$_5$ part \cite{SUGRA-KMY} is 
\begin{eqnarray}
r_{\rm Jor} = \frac{1}{\sqrt{2}}\,E_{24}\wedge (E_{22}-E_{44})\,, 
\end{eqnarray}  
where $(E_{ij})_{kl} \equiv \delta_{ik}\delta_{jl}$ and the skew-symmetrized symbol $\wedge$ is defined as  
\begin{align}
a\wedge b\equiv a\otimes b-b\otimes a\,. 
\end{align}
In fact, the associated linear R-operator exhibits the nilpotency $R_{\rm Jor}^n=0$ for $n\geq3$\,.  

\medskip 

One may adopt ``the abelian condition'' as the third property, instead of the nilpotency.    
It is easy to construct such $r$-matrices by using Cartan generators. A typical example is 
\begin{align}
r_{\rm Abe}=\sum_{\substack{i\neq j}} \mu_{ij}\, h_i\wedge h_j \,, 
\end{align}
where $\mu_{ij}=-\mu_{ji}$ are arbitrary parameters and $h_i$ are Cartan generators. 
We refer the $r$-matrices of this type as to abelian $r$-matrices because these generate abelian twists 
which are particular examples of the Drinfeld-Reshetikhin twists \cite{Drinfeld1,Drinfeld2,R}. 
These commute with each other and hence satisfy CYBE obviously. 
Note that abelian $r$-matrices are intrinsic to higher rank cases (rank $\geq 2$). 
For example, for $\alg{su}(2)$\,, these become trivial, i.e., $r_{\rm Abe}=0$\,. 
A remarkable point is that the $\kappa$-invariance and the classical integrability are ensured 
for abelian $r$-matrices, according to the observation denoted in Sec.\,\ref{2.1}.

\section{$\gamma$-deformed AdS$_5\times$S$^5$ from classical $r$-matrix}

We present here a relation between abelian 
classical $r$-matrices and $\ga$-deformed AdS$_5\times$S$^5$\,.

\subsection{$\gamma$-deformed AdS$_5\times$S$^5$ with three parameters} 

First of all, we give a brief review of gravitational duals of marginal deformations 
of the $\mathcal{N}$=4 $SU(N)$ super Yang-Mills (SYM) theory in four dimensions.  

\medskip 

For a particular class of marginal deformations of $\mathcal{N}$=4 SYM \cite{LS} called $\beta$-deformations, 
the gravitational duals were presented by Lunin and Maldacena \cite{LM}. 
Their original construction is based on an $SL(2,\mathbb{R})$ symmetry  
and a single parameter is contained. 

\medskip 

Then the solutions were generalized 
so that three parameters are contained by performing three TsT transformations \cite{Frolov}: 
1) $(\phi_1,\phi_2)_{\rm TsT}$\,, 2) $(\phi_2,\phi_3)_{\rm TsT}$ and 3) $(\phi_3,\phi_1)_{\rm TsT}$\,. 
Here $\phi_i~(i=1,2,3)$ are the Cartan directions in the S$^5$ metric 
and the symbol $(\phi_{1},\phi_2)_{\rm TsT}$\,, for example, means the following. First, 
a T-duality is performed along $\phi_1$\,. Then $\phi_2$ is shifted as $\phi_2 + \hat{\gamma}_3\,\phi_1$ 
with a constant parameter $\hat{\gamma}_3$\,. 
Finally a T-duality is taken for $\phi_1$ again.  

\medskip 

The resulting metric of three-parameter deformed AdS$_5\times$S$^5$ 
(in the string frame) and the NS-NS B-field are given by 
\begin{eqnarray}
ds^2 &=& ds^2_{\rm AdS_5} + \sum_{i=1}^3(d\rho_i^2+G \rho_i^2d\phi_i^2) 
+ G \rho_1^2\rho_2^2\rho_3^2 \left(\sum_{i=1}^3\hat{\gamma}_i\, d\phi_i\right)^2\,,   
\label{3-metric} \\ 
B_2 &=& G \left(
\hat{\gamma}_3\,\rho_1^2\rho_2^2\,d\phi_1\wedge d\phi_2 + \hat{\gamma}_1\,\rho_2^2\rho_3^2\,d\phi_2\wedge d\phi_3  
+ \hat{\gamma}_2\,\rho_3^2\rho_1^2\,d\phi_3\wedge d\phi_1 
\right)\,.
\label{3-NSNS}
\end{eqnarray}
Here there is a constraint $\sum_{i=1}^3\rho_i^2 =1$ and a scalar function $G$ is defined as  
\begin{eqnarray}
G^{-1} \equiv 1 + \hat{\gamma}_3^2\,\rho_1^2\rho_2^2 + \hat{\gamma}_1^2\,\rho_2^2\rho_3^2 
+ \hat{\gamma}_2^2\,\rho_3^2\rho_1^2\,. 
\end{eqnarray}
For the other field components, see \cite{Frolov}. This solution is often called 
the three-parameter real $\gamma$-deformed AdS$_5\times$S$^5$ background\footnote{
In \cite{LM,Frolov,B-Cherkis}, one can see the classical potential in the corresponding 
deformed $\mathcal{N}=4$ SYM theory. 
This potential gets quantum corrections and conformal invariance is broken at quantum level \cite{HU}. 
For this point, we would like to thank J.~Fokken, C.~Sieg and M.~Wilhelm, and D.~Giataganas. }.  
When $\hat{\gamma}_1=\hat{\gamma}_2=\hat{\gamma}_3 \equiv \hat{\gamma}$\,, 
the original Lunin-Maldacena background for the real $\beta$-deformation is reproduced. 

\medskip

In Sec.\,\ref{3.2}, we will present a classical $r$-matrix corresponding to one of the TsT transformations 
used above.

\subsection{One-parameter case \label{3.2}} 

As a warm-up, let us consider a simple example of classical $r$-matrix, 
\begin{eqnarray}
r_{\rm Abe}^{(\mu)} = \mu\, h_1 \wedge h_2\,. \label{1-para}
\end{eqnarray}
Here $\mu$ is a deformation parameter and the fundamental 
representation of Cartan generators of $\mathfrak{su}(4)$\,, $h_1$ and $h_2$ are defined as 
\begin{align}
h_1 \equiv {\rm diag}(-1,1,-1,1)\,,\qquad 
h_2 \equiv {\rm diag}(-1,1,1-,1)\,.
\label{h1h2}
\end{align}
The action of the associated linear $R_{\rm Abe}^{(\mu)} $ operator is given by  
\begin{eqnarray}
R_{\rm Abe}^{(\mu)}(h_1) = -\mu\, h_2\,, \qquad
R_{\rm Abe}^{(\mu)}(h_2) = \mu\, h_1\,, \qquad
R_{\rm Abe}^{(\mu)}(\text{other}) =0\,, 
\end{eqnarray}
and hence only the S$^5$ part of AdS$_5\times$S$^5$ is deformed. 

\medskip 

Since we are interested in deformations of S$^5$\,,   
it is convenient to restrict the current $A_\al\in \alg{su}(2,2|4)$ 
to the $\alg{su}(4)$ subalgebra as follows:  
\begin{align}
A_\al=g^{-1}\partial_\al g 
\qquad \text{with} \qquad 
g \in SU(4)/SO(5)\,. 
\end{align}
With this setup, the S$^5$ part of the classical action \eqref{action} is reduced to  
\begin{align}
&S=\int^\infty_{-\infty}\!\!\!d\tau\int^{2\pi}_0\!\!\!d\sigma~(L_G+L_B) \,, 
\label{S5action} \\
&L_G=+\frac{1}{2}\ga^{\al\be}
{\rm Tr}\left[A_\al P_2\circ\frac{1}{1-2\eta \left[R_{\rm Abe}^{(\mu)}\right]_g\circ P_2}A_\be\right]\,,  \el\\
&L_B=-\frac{1}{2}\ep^{\al\be} 
{\rm Tr}\left[A_\al P_2\circ\frac{1}{1-2\eta \left[R_{\rm Abe}^{(\mu)}\right]_g\circ P_2}A_\be\right]\,,  \el
\end{align}
where $L_G$ is the sigma model part and $L_B$ represents the coupling to the NS-NS two-form. 

\medskip 

Then the classical Lagrangian given in \eqref{S5action} can be rewritten as 
\begin{align}
L_G&=-\frac{\ga^{\al\be}}{2}\Bigl[\partial_\al r\partial_\be r
+\sin^2r \partial_\al \ze\partial_\be \ze 
+\cos^2r \partial_\al \phi_3\partial_\be \phi_3  
\el\\ 
&\qquad \qquad 
+\frac{\sin^2r }{1+16\eta^2\mu^2\sin^4r\sin^22\ze}
\left(\cos^2\ze\partial_\al \phi_1\partial_\be \phi_1
+\sin^2\ze\partial_\al \phi_2\partial_\be \phi_2\right) 
\Bigr] \,, 
\label{G1param} \\
L_B&=\frac{2 \eta\mu\sin^4r\sin^22\ze }{1 + 16 \eta^2\mu^2\sin^4r\sin^22\ze} 
\ep^{\al\be} \partial_\al \phi_1\partial_\be \phi_2 \,.  
\label{B1param} 
\end{align}
For the derivation, see Appendix C. 

\medskip 

By imposing a parameter relation,  
\begin{align}
0=\hat\ga_1=\hat\ga_2\,, \qquad  
8\eta\mu=\hat\ga_3 \,, 
\end{align}
and performing the following coordinate transformation,    
\begin{align}
\rho_1=\sin r\cos\zeta\,, \qquad 
\rho_2=\sin r\sin\zeta \,, \qquad
\rho_3=\cos r\,,     
\label{coord-trf}
\end{align}
we find that the Lagrangian \eqref{G1param} and \eqref{B1param} are nothing but 
the ones obtained from the $\ga$-deformed metric \eqref{3-metric} 
and the NS-NS two-form \eqref{3-NSNS}, respectively.  
Thus we have shown that the classical $r$-matrix (\ref{1-para}) corresponds to 
a TsT transformation $(\phi_1,\phi_2)_{\rm TsT}$\,. 

\medskip 

It would be interesting to reinterpret this result from the viewpoint of a twisted boundary condition   
by following \cite{AAF}. In particular, there should be some relation between the classical $r$-matrix 
and the boundary condition.

\subsection{Three-parameter case}

Now it would be easy to deduce the classical $r$-matrix that corresponds to the three-parameter 
deformed solution, according to the result obtained in the previous subsection. The candidate $r$-matrix is represented by 
\begin{eqnarray}
r_{\rm Abe}^{(\mu_1,\mu_2,\mu_3)} = \mu_3\, h_1 \wedge h_2 + \mu_1\, h_2 \wedge h_3 + \mu_2\, h_3 \wedge h_1\,,
\label{3-para}
\end{eqnarray}
where $\mu_i$ and $h_i$ $(i=1,2,3)$ are deformation parameters and the Cartan generators of $\mathfrak{su}(4)$\,. 
For $h_1$ and $h_2$\,, see \eqref{h1h2}. The remaining $h_3$ is defined as  
\begin{align}
h_3 \equiv {\rm diag}(1,1,-1,-1)\,. 
\end{align}
Then the action of the associated linear $R_{\rm Abe}^{(\mu_1,\mu_2,\mu_3)}$ is given by  
\begin{align}
&R_{\rm Abe}^{(\mu_1,\mu_2,\mu_3)}(h_1) = \mu_2\,h_3 - \mu_3\,h_2\,,
\qquad 
R_{\rm Abe}^{(\mu_1,\mu_2,\mu_3)}(h_2) = \mu_3\,h_1 - \mu_1\,h_3\,,
\el \\
&R_{\rm Abe}^{(\mu_1,\mu_2,\mu_3)}(h_3) = \mu_1\,h_2 - \mu_2\,h_1\,,
\qquad 
R_{\rm Abe}^{(\mu_1,\mu_2,\mu_3)}(\text{other}) =0\,. 
\label{3-R}
\end{align}

\medskip 

With the following parameter identification, 
\begin{align}
8\eta\,\mu_1=\hat\ga_1\,, \quad 
8\eta\,\mu_2=\hat\ga_2\,, \quad 
8\eta\,\mu_3=\hat\ga_3\,, 
\end{align}
the deformed Lagrangians $L_G$ and $L_B$ turn out to be 
\begin{align}
L_G &= -\frac{\ga^{\al\be}}{2} \Bigl[\sin^2 r\, \partial_\al r\partial_\be r  
+(\cos r \sin \ze \partial_\al r+\sin r\cos \ze \partial_\al \ze)
(\cos r \sin \ze \partial_\be r+\sin r\cos \ze \partial_\be \ze) \el\\
&\qquad\quad 
+(\cos r \cos \ze\partial_\al r-\sin r\sin \ze \partial_\al \ze)
(\cos r \cos \ze\partial_\be r-\sin r\sin \ze \partial_\be \ze)  \el \\
&\qquad\quad 
+G\bigl[\sin^2 r(\cos^2 \ze \partial_\al \phi_1\partial_\be \phi_1
+\sin^2 \ze \partial_\al \phi_2\partial_\be \phi_2)
+\cos^2r \partial_\al \phi_3\partial_\be \phi_3 
\el \\
&\qquad\qquad \quad 
+\cos^2 r \sin^4 r  \cos^2 \ze\sin^2 \ze 
\bigl(\textstyle{\sum_{i}}\hat\ga_i \partial_\al \phi_i\bigr) 
\bigl(\textstyle{\sum_{j}}\hat\ga_j \partial_\be \phi_j\bigr) \bigr] 
\Bigr]\,, 
\label{3.15} \\
L_B&=\ep^{\al\be} G \Bigl[
\hat\ga_3\sin^4r \sin^2\ze\cos^2\ze \partial_\al \phi_1\partial_\be \phi_2  
\el\\
&\qquad\qquad
+\sin^2 r\cos^2r\left(\hat\ga_1\sin^2\ze \partial_\al \phi_2\partial_\be \phi_3 
+\hat\ga_2\cos^2\ze \partial_\al \phi_3\partial_\be \phi_1 \right)
\Bigr]\,. \label{3.16}
\end{align}
where the function $G$ is rewritten as  
\begin{align}
G^{-1}&=1+\cos^2 r \sin^2 r\,(\hat\ga_1^2\sin^2 \ze+\hat\ga_2^2\cos^2 \ze )
+\hat\ga_3^2\sin^4r\cos^2\ze\sin^2\ze \,.  
\end{align}
Finally, with the coordinate transformation \eqref{coord-trf}\,, 
the deformed metric and NS-NS two-form obtained from (\ref{3.15}) and (\ref{3.16})
exactly agree with the three parameter $\ga$-deformed metric 
\eqref{3-metric} and \eqref{3-NSNS}, respectively. The derivation of them is described in detail in Appendix C. 
Thus the classical $r$-matrix \eqref{3-para} corresponds to three TsT transformations 
$(\phi_1,\phi_2)_{\rm TsT}$\,, $(\phi_2,\phi_3)_{\rm TsT}$ and $(\phi_3,\phi_1)_{\rm TsT}$\,. 

\medskip 

The gravity dual for the real $\beta$-deformation is realized as a particular case 
with $\hat{\gamma}_1=\hat{\gamma}_2=\hat{\gamma}_3=\hat{\gamma}$\,. 
It is wroth noting that complex $\beta$-deformations are argued to yield non-integrable 
backgrounds \cite{DG}. Probably, there would be no classical $r$-matrix for the complex $\beta$-deformations 
within the class that allows the Lax pair construction. 
It may be intriguing to look for the corresponding classical $r$-matrix by admitting that the integrability is broken.

\medskip 

Now the relation between classical $r$-matrices and the $\gamma$-deformed geometries has been clarified. 
For the $\gamma$-deformed geometries, various things are understood such as the deformed potential in $\mathcal{N}$=4 
SYM \cite{LM,Frolov,B-Cherkis}, the twisted Bethe ansatz \cite{Beauty,ABBN} and the world-sheet S-matrix \cite{AKL}. 
It would be interesting to argue the relation between them and the classical $r$-matrices used 
in the Yang-Baxter sigma model approach.

\section{Conclusion and discussion}

In this note, we have considered $\gamma$-deformations of the AdS$_5\times$S$^5$ superstring 
as Yang-Baxter sigma models with classical $r$-matrices satisfying CYBE. 
An essential point is that the classical $r$-matrices are composed of Cartan generators only 
and generate abelian twists. 
They do not satisfy the nilpotency condition in comparison to Jordanian deformations 
considered in \cite{KMY-Jordanian-typeIIB}.  
We have presented examples of the $r$-matrices that lead to real $\gamma$-deformations of 
the AdS$_5\times$S$^5$ superstring. 

\medskip 

Based on our result, one may expect that TsT transformed AdS$_5\times$S$^5$ geometries 
could be classified in terms of classical $r$-matrices satisfying CYBE. 
The conjectured relations are summarized in Tab.\,\ref{table1}, 
though it is still necessary to make efforts to get supporting evidence.    
A support is that the type IIB supergravity solution 
constructed with a Jordanian twist \cite{SUGRA-KMY} may be regarded 
as a null Melvin twist, basically following the argument in Appendix C of \cite{HRR}. 
We will report on the details in the near future \cite{future}. 
There are many gravitational solutions obtained as 
TsT transformed or null Melvin twisted AdS$_5\times$S$^5$\,. 
There should be a classical $r$-matrix for each of them. 

\medskip 

At the beginning, the Yang-Baxter sigma model approach has been regarded as a prescription for 
standard $q$-deformations. Now it seems likely that it potentially contains 
much broader applications to study integrable deformations. 
It would provide a guiding principle for classifying possible integrable deformations and the corresponding 
gravity solutions in terms of solutions of CYBE, 
which should be called the gravity/CYBE correspondence.   
 
\begin{table}[t]
\begin{center}
\begin{tabular}{l|l|l} 
\hline 
 Operations in SUGRA & Integrable deformations & Classical $r$-matrices  \\ 
\hline\hline  
  TsT transformations & Abelian twists & CYBE, skew-symm., abelian \\ 
  Null Melvin twists & Jordanian twists & CYBE, skew-symm., nilpotent \\ 
  \hline 
\end{tabular}
\end{center}
\vspace*{-0.5cm}
\caption{\footnotesize Relations among operations in SUGRA, integrable deformations and 
classical $r$-matrices. \label{table1}}
\end{table}

\subsection*{Acknowledgments}

We would like to thank Io Kawaguchi for useful discussions and collaborations at the earlier stage. 
We are also grateful to thank Sanefumi Moriyama, Ryo Suzuki and Masato Taki for useful discussions. 
T.M.\ also thanks Gleb Arutyunov and Riccardo Borsato for useful discussions.  
T.M.\ is supported by the Netherlands Organization for Scientific 
Research (NWO) under the VICI grant 680-47-602.  
T.M.'s work is also part of the ERC Advanced grant research programme 
No.~246974, ``Supersymmetry: a window to non-perturbative physics" 
and of the D-ITP consortium, a program of the NWO that is funded by the 
Dutch Ministry of Education, Culture and Science (OCW).

\appendix

\section*{Appendix} 

\section{Our notation and convention}

Our notation is summarized here. We basically follow the one used in \cite{AF-review}\,. 

\medskip 

An element of $\mathfrak{su}(2,2|4)$ is identified with 
an $8\times 8$ supermatrix: 
\begin{eqnarray}
M=\begin{bmatrix}
~m~&~\xi~\\
~\zeta~&~n~
\end{bmatrix}\,. 
\end{eqnarray}
Here $m$ and $n$ are $4\times 4$ matrices with Grassmann even elements, 
while $\xi$ and $\zeta$ are $4\times 4$ matrices with Grassmann odd elements. 
These matrices satisfy a reality condition. Then $m$ and $n$ belong to $\mathfrak{su}(2,2)=\mathfrak{so}(2,4)$ and  
$\mathfrak{su}(4)=\mathfrak{so}(6)$\,, respectively.  

\medskip 

In this note we are concerned with deformations of the S$^5$ part. 
Hence it is helpful to prepare an explicit basis of $\mathfrak{su}(4)$\,.   

\medskip 

Let us first introduce the following $\gamma$ matrices: 
\begin{eqnarray}
&&\gamma_1=
\begin{bmatrix}
~0~&~0~&~0~&-1\\
~0~&~0~&~1~&~0\\
~0~&~1~&~0~&~0\\
-1~&~0~&~0~&~0\\
\end{bmatrix}\,, \quad
\gamma_2=
\begin{bmatrix}
~0~&~0~&~0~&~i~\\
~0~&~0~&~i~&~0~\\
~0~&-i~&~0~&~0~\\
-i~&~0~&~0~&~0~\\
\end{bmatrix}\,, \quad 
\gamma_3=
\begin{bmatrix}
~0~&~0~&~1~&~0~\\
~0~&~0~&~0~&~1~\\
~1~&~0~&~0~&~0~\\
~0~&~1~&~0~&~0~\\
\end{bmatrix}\,, \nonumber \\
&&\gamma_4=
\begin{bmatrix}
~0~&~0~&-i~&~0~\\
~0~&~0~&~0~&~i~\\
~i~&~0~&~0~&~0~\\
~0~&-i~&~0~&~0~\\
\end{bmatrix}\,, \quad
\gamma_5=-\gamma_1\gamma_2\gamma_3\gamma_4=
\begin{bmatrix}
~1~&~0~&~0~&~0~\\
~0~&~1~&~0~&~0~\\
~0~&~0~&-1~&~0~\\
~0~&~0~&~0~&-1~\\
\end{bmatrix}\,. 
\end{eqnarray}
It is easy to see that
\begin{eqnarray}
n_{ij}=\frac{1}{4}\left[\gamma_i,\gamma_j\right] \qquad (i,j=1,\ldots,5)  
\end{eqnarray}
generate $\mathfrak{so}(5)$ by using the Clifford algebra 
\begin{eqnarray}
\left\{\gamma_i,\gamma_j\right\}=2\delta_{ij}\,. 
\end{eqnarray}
Note that $\mathfrak{so}(6)$ is spanned by the set of the generators,
\begin{align}
n_{ij}\,, \qquad n_{i6}=-n_{6i}=\frac{i}{2}\gamma_i\,. 
\end{align}

\section{Rewriting $\gamma$-deformed backgrounds}

For our purpose, it is convenient to rewrite the metric \eqref{3-metric}
and the NS-NS B-field \eqref{3-NSNS} in terms of angle variables;  
\begin{align}
\rho_1=\sin r\cos\zeta\,, \qquad 
\rho_2=\sin r\sin\zeta \,, \qquad
\rho_3=\cos r\,.    
\end{align}
With the above coordinates, the metric and the NS-NS two-form are given by 
\begin{align}
ds^2 &= ds^2_{\rm AdS_5} +\sin^2 r\, dr^2 
+(\cos r \sin \ze\, dr + \sin r\cos \ze\, d\ze)^2
+(\cos r \cos \ze\, dr - \sin r\sin \ze\, d\ze)^2 \el \\
&\quad + G\bigl[\sin^2 r\,(\cos^2 \ze\, d\phi_1^2+\sin^2 \ze\, d\phi_2^2)+\cos^2r\, d\phi_3^2 \el\\
&\qquad\quad  +\cos^2 r \sin^4 r  \cos^2 \ze\sin^2 \ze \,\bigl(\textstyle{\sum_{i}}\hat\ga_i d\phi_i\bigr)^2 \bigr]\,,  
\label{gmetric}
\\
B_2&=
G\bigl[\hat\ga_3\sin^4r \sin^2\ze\cos^2\ze d\phi_1\wedge d\phi_2
\el \\
&\qquad 
+\sin^2 r\cos^2r\left(\hat\ga_1\sin^2\ze d\phi_2\wedge d\phi_3 
+\hat\ga_2\cos^2\ze d\phi_3\wedge d\phi_1 \right) \bigr]\,,  
\label{gNSNS}
\end{align}
where the scalar function $G$ is also rewritten as 
\begin{align}
G^{-1}&=1+\cos^2 r \sin^2 r\,(\hat\ga_1^2\sin^2 \ze+\hat\ga_2^2\cos^2 \ze )
+\hat\ga_3^2\sin^4r\cos^2\ze\sin^2\ze \,.  
\end{align}

\section{Derivation of deformed actions}

Here we describe in detail the derivation of the deformed action with the classical $r$-matrix (\ref{3-para}). 
The AdS$_5$ part is not deformed and hence we will concentrate on the S$^5$ part hereafter. 

\medskip 

Let us adopt the following coset parametrization \cite{ABF}\,:   
\begin{eqnarray}
g =\Lambda(\phi_1,\phi_2,\phi_3)\, \Xi(\ze)\, \check g_{\rm r}(r)    
\quad \in~ SU(4)/SO(5) 
\end{eqnarray}
with the matrices $\Lambda, \Xi$ and $\check g_{\rm r}$ defined as 
\begin{eqnarray}
&& \Lambda(\phi_1,\phi_2,\phi_3) = \exp\left[\frac{i}{2}(\phi_1h_1+\phi_2h_2+\phi_3h_3)\right]\,, \el\\
&& \hspace*{-1cm}
\Xi(\ze) = 
\begin{pmatrix} 
\cos\frac{\ze}{2} & \sin\frac{\ze}{2}& 0& 0 \\ 
-\sin\frac{\ze}{2} & \cos\frac{\ze}{2}& 0& 0 \\ 
0& 0&\cos\frac{\ze}{2} & -\sin\frac{\ze}{2}  \\ 
0& 0&\sin\frac{\ze}{2} & \cos\frac{\ze}{2} 
\end{pmatrix}\,, 
\quad 
\check g_{\rm r}(r) 
=\begin{pmatrix} 
\cos\frac{r}{2} & 0& 0& i\sin\frac{r}{2}  \\ 
0 & \cos\frac{r}{2} & -i\sin\frac{r}{2} & 0 \\ 
0& -i\sin\frac{r}{2} &\cos\frac{r}{2}  & 0  \\ 
i\sin\frac{r}{2} & 0&0 & \cos\frac{r}{2}  
\end{pmatrix}\,,     
\end{eqnarray}
where $h_i~(i=1,2,3)$ are diagonal matrices given by 
\begin{align}
h_1={\rm diag}(-1,1,-1,1)\,,\quad 
h_2={\rm diag}(-1,1,1-,1)\,,\quad 
h_3={\rm diag}(1,1,-1,-1)\,. 
\end{align}
These correspond to the Cartan generators of $\mathfrak{su}(4)$\,. 

\medskip 

With this parametrization, the S$^5$ part of the Lagrangian \eqref{action} can be rewritten as 
\begin{align}
L &=\frac{1}{2} 
(\gamma^{\alpha\beta}-\epsilon^{\alpha\beta}){\rm Tr}\left[
A_\alpha P_2\circ
\frac{1}{1-2\eta\bigl[R_{\rm Abe}^{(\mu_1,\mu_2,\mu_3)}\bigr]_g\circ P_2}A_\beta\right]\,,  
\end{align}
where $A_\al=g^{-1}\partial_\al g$ is restricted to $\alg{su}(4)$ and the R-operator  
is defined in (\ref{3-R}). 
For later argument, 
it is convenient to divide the Lagrangian $L$ into the two parts like $L=L_G+L_B$\,, 
where $L_G$ is the metric part and $L_B$ is the coupling to the NS-NS two-form, respectively:  
\begin{eqnarray}
L_G &\equiv& -\frac{1}{2} [\Tr(A_{\tau}P_2(J_{\tau}))-\Tr(A_{\sigma}P_2(J_{\sigma}))] \,, \nonumber \\ 
L_B &\equiv& -\frac{1}{2} [\Tr(A_{\tau}P_2(J_{\sigma}))-\Tr(A_{\sigma}P_2(J_{\tau}))] \,.  
\end{eqnarray}
Here the deformed current $J_\al$ is defined as 
\begin{align}
J_\al\equiv \frac{1}{1-2\eta\bigl[R_{\rm Abe}^{(\mu_1,\mu_2,\mu_3)}\bigr]_g\circ P_2}A_\al\,. 
\end{align}
This current contains $\mu_i~(i=1,2,3)$ and the normalization factor $\eta$\,. 

\medskip 

To derive the explicit form of the deformed Lagrangian, 
it is sufficient to compute the projected current $P_2(J_\al)$ 
rather than $J_\al$ itself.  
Hence the problem is boiled down to solving the following equation,  
\begin{align}
\left(1-2\eta P_2\circ \left[R_{\rm Abe}^{(\mu_1,\mu_2,\mu_3)}\right]_g\right)P_2(J_\al)=P_2(A_\al)\,. 
\label{equation}
\end{align}
Note that $P_2(A_\alpha)$ is expanded with gamma matrices $\ga_i$ as follows:    
\begin{align}
P_2(A_\al)&=-\frac{i}{2}\bigl(
 \ga_1\partial_\al r+ 
\sin r\,(\ga_2\cos\zeta\, \partial_\al \phi_1 + \ga_3\partial_\al \zeta + \ga_4\sin\zeta\, \partial_\al \phi_2) 
-\ga_5\cos r\, \partial_\al \phi_3 \bigr)\,. 
\label{P_2(A)} 
\end{align}
Then, by combining the expression (\ref{P_2(A)}) with \eqref{equation}\,, 
the projected deformed current $P_2(J_\alpha)$ can be obtained as 
\begin{align}
P_2(J_\alpha)=\gamma_1\,j_\alpha^1+\gamma_2\,j_\alpha^2
+\gamma_3\,j_\alpha^3+\gamma_4\,j_\alpha^4+\gamma_5\,j_\alpha^5\,, 
\end{align}
with the coefficients 
\begin{align}
&j_\alpha^1=-\frac{i}{2}\partial_\al r \el \\
&j_\alpha^2=-\frac{i}{2}
\frac{\sin r \cos\ze}{1 + 16 \eta^2 [\mu_1^2 \sin^22\ze \sin^4r  +
(\mu_2^2 \sin^2\ze+\mu_3^2 \cos^2\ze)\sin^22r  ]} \el \\
&\qquad \times \Bigl[
(1 + 16 \eta^2 \mu_2^2 \sin^22r \sin^2\ze)
\partial_\al \phi_1 \el \\
&\qquad \quad 
+8 \eta (\mu_1+8 \eta\mu_2\mu_3 \cos^2r)\sin^2r \sin^2\ze 
\partial_\al \phi_2 \el \\
&\qquad \quad 
+ 8 \eta (-\mu_3 + 8\eta\mu_1\mu_2 \sin^2r \sin^2\ze) \cos^2r 
\partial_\al \phi_3 \Bigr]
\,, \nonumber \\
&j_\alpha^3=-\frac{i}{2}\partial_\al\zeta \sin r 
\el \\
&j_\alpha^4=-\frac{i}{2}\frac{\sin r \sin\ze}{1 + 16 \eta^2 [\mu_1^2 \sin^22\ze \sin^4r 
+(\mu_2^2 \sin^2\ze+\mu_3^2 \cos^2\ze)\sin^22r  ]} \el \\
&\qquad 
\times \Bigl[
(1+16 \eta^2\mu_3^2 \sin^22r \cos^2\ze) 
\partial_\al \phi_2 \el \\
&\qquad \quad 
+8 \eta (-\mu_1 + 8 \eta\mu_2\mu_3  \cos^2r)  \sin^2r \cos^2\ze 
\partial_\al \phi_1 
\el \\
&\qquad \quad 
+8 \eta (\mu_2+8\eta\mu_1\mu_3\cos^2\ze \sin^2r)\cos^2r 
\partial_\al \phi_3
\Bigr]\,,  
\el \\ 
&j_\al^5=\frac{i}{2} \frac{\cos r }{1 + 16 \eta^2 [\mu_1^2 \sin^22\ze \sin^4r 
+(\mu_2^2 \sin^2\ze+\mu_3^2 \cos^2\ze)\sin^22r  ]} \el \\
&\qquad \times \Bigl[
(1+16 \eta^2\mu_1^2 \sin^4r\sin^22\ze) 
\partial_\al \phi_3
\el \\
&\qquad \quad 
+8\eta (\mu_3 + 8\eta\mu_1\mu_2 \sin^2r\sin^2\ze) \sin^2r\cos^2\ze  
\partial_\al \phi_1
\el\\
&\qquad \quad 
+ 8 \eta(-\mu_2+8\eta\mu_1\mu_3 \cos^2\ze\sin^2r ) \sin^2r\sin^2\ze 
\partial_\al \phi_2
\Bigr]  \,. 
\end{align}
Finally, $L_G$ and $L_B$ are given by  
\begin{align}
L_G &= -\frac{\ga^{\al\be}}{2} \Bigl[\sin^2 r\, \partial_\al r\partial_\be r  
+(\cos r \sin \ze \partial_\al r+\sin r\cos \ze \partial_\al \ze)
(\cos r \sin \ze \partial_\be r+\sin r\cos \ze \partial_\be \ze) \el\\
&\qquad\quad  
+(\cos r \cos \ze\partial_\al r-\sin r\sin \ze \partial_\al \ze)
(\cos r \cos \ze\partial_\be r-\sin r\sin \ze \partial_\be \ze)  \el \\
&\qquad\quad 
+G\bigl[\sin^2 r(\cos^2 \ze \partial_\al \phi_1\partial_\be \phi_1
+\sin^2 \ze \partial_\al \phi_2\partial_\be \phi_2)
+\cos^2r \partial_\al \phi_3\partial_\be \phi_3 \el \\
&\qquad\qquad \quad 
+\cos^2 r \sin^4 r  \cos^2 \ze\sin^2 \ze 
\bigl(\textstyle{\sum_{i}}\hat\ga_i \partial_\al \phi_i\bigr) 
\bigl(\textstyle{\sum_{j}}\hat\ga_j \partial_\be \phi_j\bigr) \bigr] 
\Bigr]\,,  
\label{LG3}
\\
L_B&=\ep^{\al\be} G \Bigl[
\hat\ga_3\sin^4r \sin^2\ze\cos^2\ze\, \partial_\al \phi_1\,\partial_\be \phi_2  
\el\\
&\qquad\qquad
+\sin^2 r\cos^2r\left(\hat\ga_1\sin^2\ze\, \partial_\al \phi_2\partial_\be \phi_3 
+\hat\ga_2\cos^2\ze \partial_\al \phi_3\,\partial_\be \phi_1 \right) 
\Bigr]\,.
\label{LB3}
\end{align}
Here the deformation parameters $\mu_i$~($i=1,2,3$) in the classical $r$-matrix (\ref{3-para}) 
are identified with those of $\ga$-deformations $\hat{\gamma}_i~(i=1,2,3)$ through the relations 
\begin{align} 
8\eta\,\mu_i = \hat\ga_i \qquad (i=1,2,3)\,. 
\end{align}
Now one can derive the metric and the NS-NS two-form from \eqref{LG3} and \eqref{LB3}\,. 
The resulting metric and two-form exactly agree with the metric \eqref{gmetric} and two-form \eqref{gNSNS} 
for the three-parameter $\ga$-deformed S$^5$\,. 

\medskip 

As a particular case, the one-parameter deformed Lagrangian given by \eqref{G1param} 
and \eqref{B1param} is given by taking the following parameters:  
\begin{align}
\mu_1=\mu_2=0 \qquad\text{and}\qquad  \mu_3=\mu\,.   
\end{align}


\begin{thebibliography}{99}

\bibitem{M}  
  J.~M.~Maldacena,
  ``The large N limit of superconformal field theories and supergravity,''
  Adv.\ Theor.\ Math.\ Phys.\  {\bf 2} (1998) 231
  [Int.\ J.\ Theor.\ Phys.\  {\bf 38} (1999) 1113]. 
  [arXiv:hep-th/9711200].

\bibitem{GKP}
S.~S.~Gubser, I.~R.~Klebanov and A.~M.~Polyakov,
``Gauge theory correlators from non-critical string theory,''
Phys.\ Lett.\ B {\bf 428} (1998) 105 [arXiv:hep-th/9802109]. 

\bibitem{W}
E.~Witten, 
``Anti-de Sitter space and holography,''
Adv.\ Theor.\ Math.\ Phys.\  {\bf 2} (1998) 253 [arXiv:hep-th/9802150].

\bibitem{review}
  N.~Beisert {\it et al.},
  ``Review of AdS/CFT Integrability: An Overview,'' 
  Lett.\ Math.\ Phys.\ {\bf 99} (2012) 3 [arXiv:1012.3982 [hep-th]]. 

\bibitem{MT}
  R.~R.~Metsaev and A.~A.~Tseytlin,
  ``Type IIB superstring action in AdS$_5\times$S$^5$ background,''  
  Nucl.\ Phys.\ B {\bf 533} (1998) 109  [hep-th/9805028].  

\bibitem{BPR}
  I.~Bena, J.~Polchinski and R.~Roiban,
  ``Hidden symmetries of the AdS$_5\times$S$^5$ superstring,''
  Phys.\ Rev.\ D {\bf 69} (2004) 046002
  [hep-th/0305116].
  
\bibitem{RS}  
 R.~Roiban and W.~Siegel,
  ``Superstrings on AdS$_5\times$S$^5$ supertwistor space,''  
JHEP {\bf 0011} (2000) 024  [hep-th/0010104].  
  
\bibitem{Hatsuda}
  M.~Hatsuda and K.~Yoshida,
  ``Classical integrability and super Yangian of superstring on AdS$_5\times$S$^5$,''  
Adv.\ Theor.\ Math.\ Phys.\  {\bf 9} (2005) 703  [hep-th/0407044];   
  ``Super Yangian of superstring on AdS$_5\times$S$^5$ revisited,''  
Adv.\ Theor.\ Math.\ Phys.\  {\bf 15} (2011) 1485  [arXiv:1107.4673 [hep-th]].

\bibitem{Zarembo-symmetric}
  K.~Zarembo,
  ``Strings on Semisymmetric Superspaces,''
  JHEP {\bf 1005} (2010) 002
  [arXiv:1003.0465 [hep-th]]. 

\bibitem{Wulf}
  L.~Wulff,
  ``Superisometries and integrability of superstrings,''  arXiv:1402.3122 [hep-th]. 
  
%%%%%%%%%%%%%%%%%%%%%%%%%%%%%%%%%%%%%%%%%%%%%%%%%%%%%%%%%%%%%  
  
\bibitem{Drinfeld1}
  V.~G.~Drinfel'd,
  ``Hopf algebras and the quantum Yang-Baxter equation,'' 
  Sov.\ Math.\ Dokl.\ {\bf 32} (1985) 254.  

\bibitem{Drinfeld2}
 V.~G.~Drinfel'd,
  ``Quantum groups,''
  J.\ Sov.\ Math.\  {\bf 41} (1988) 898 
  [Zap.\ Nauchn.\ Semin.\  {\bf 155}, 18 (1986)].

\bibitem{Jimbo}
  M.~Jimbo,
  ``A $q$ difference analog of $U(g)$ and the Yang-Baxter equation,''
  Lett.\ Math.\ Phys.\  {\bf 10} (1985) 63.      

\bibitem{Jordanian}
  A.~Stolin and P.~P.~Kulish, 
  ``New rational solutions of Yang-Baxter equation and deformed Yangians,''
  Czech.\ J.\ Phys.\ {\bf 47} (1997) 123 [arXiv:q-alg/9608011].  

\bibitem{KLM}
  P.~P.~Kulish, V.~D.~Lyakhovsky and A.~I.~Mudrov,
  ``Extended jordanian twists for Lie algebras,''  
  J.\ Math.\ Phys.\  {\bf 40} (1999) 4569  [math/9806014 [math.QA]].      


%%%%%%%%%%%%%%%%%%%%%%%%%%%%%%%%%%%%%%%%%%%%%%%%%%%%%%%%%
 
\bibitem{Klimcik}
 C.~Klimcik,
  ``Yang-Baxter sigma models and dS/AdS T duality,''  
JHEP {\bf 0212} (2002) 051  [hep-th/0210095]; 
  ``On integrability of the Yang-Baxter sigma-model,''  
J.\ Math.\ Phys.\  {\bf 50} (2009) 043508  [arXiv:0802.3518 [hep-th]]; 
 ``Integrability of the bi-Yang-Baxter sigma model,'' arXiv:1402.2105 [math-ph].  

\bibitem{Squellari}
  R.~Squellari,
  ``Yang-Baxter $\sigma$ model: Quantum aspects,''  arXiv:1401.3197 [hep-th]. 

%%%%%%%%%%%%%%%%%%%%%%%%%%%%%%%%%%%%%%%%%%%%%%%%%%%%%%%%% 

\bibitem{SYY}
  S.~Schafer-Nameki, M.~Yamazaki and K.~Yoshida,
  ``Coset Construction for Duals of Non-relativistic CFTs,''
  JHEP {\bf 0905} (2009) 038
  [arXiv:0903.4245 [hep-th]].         
 
%%%%%%%%%%%%%%%%%%%%%%%%%%%%%%%%%%%%%%  
  
\bibitem{Cherednik}
  I.~V.~Cherednik, 
  ``Relativistically Invariant Quasiclassical Limits Of Integrable
  Two-Dimensional Quantum Models,''
  Theor.\ Math.\ Phys.\  {\bf 47} (1981) 422
  [Teor.\ Mat.\ Fiz.\  {\bf 47} (1981) 225].

\bibitem{FR}
  L.~D.~Faddeev and N.~Y.~Reshetikhin,
  ``Integrability of the principal chiral field model in (1+1)-dimension,''
  Annals Phys.\  {\bf 167} (1986) 227.      

\bibitem{BFP}
  J.~Balog, P.~Forgacs and L.~Palla,
  ``A two-dimensional integrable axionic sigma model and T duality,''  
  Phys.\ Lett.\ B {\bf 484} (2000) 367  
  [hep-th/0004180].  

\bibitem{Mohammedi}
  N.~Mohammedi,
  ``On the geometry of classically integrable two-dimensional non-linear sigma models,''
  Nucl.\ Phys.\ B {\bf 839} (2010) 420
  [arXiv:0806.0550 [hep-th]].  

\bibitem{KY}
  I.~Kawaguchi and K.~Yoshida,
  ``Hidden Yangian symmetry in sigma model on squashed sphere,''
  JHEP {\bf 1011} (2010) 032. 
  [arXiv:1008.0776 [hep-th]]. 

\bibitem{KYhybrid}
  I.~Kawaguchi and K.~Yoshida,
  ``Hybrid classical integrability in squashed sigma models,''
  Phys.\ Lett.\ B\ {\bf 705} (2011) 251
  [arXiv:1107.3662 [hep-th]]; 
   ``Hybrid classical integrable structure of squashed sigma models: A short summary,''  
  J.\ Phys.\ Conf.\ Ser.\  {\bf 343} (2012) 012055 
  [arXiv:1110.6748 [hep-th]].    
  
\bibitem{KMY-QAA}
  I.~Kawaguchi, T.~Matsumoto and K.~Yoshida,
  ``The classical origin of quantum affine algebra in squashed sigma models,''  
  JHEP {\bf 1204} (2012) 115  [arXiv:1201.3058 [hep-th]].  

\bibitem{KMY-monodromy}
  I.~Kawaguchi, T.~Matsumoto and K.~Yoshida,
  ``On the classical equivalence of monodromy matrices in squashed sigma model,''  
  JHEP {\bf 1206} (2012) 082  [arXiv:1203.3400 [hep-th]].

\bibitem{ORU}
  D.~Orlando, S.~Reffert and L.~I.~Uruchurtu,
  ``Classical integrability of the squashed three-sphere, warped AdS3 and
  Schr$\ddot{\rm o}$dinger spacetime via T-Duality,''
  J.\ Phys.\ A  {\bf 44} (2011) 115401.
  [arXiv:1011.1771 [hep-th]]. 
  
\bibitem{KOY}
  I.~Kawaguchi, D.~Orlando and K.~Yoshida,
  ``Yangian symmetry in deformed WZNW models on squashed spheres,''
  Phys.\ Lett.\  B {\bf 701} (2011) 475. 
  [arXiv:1104.0738 [hep-th]]; 
  I.~Kawaguchi and K.~Yoshida,
  ``A deformation of quantum affine algebra in squashed WZNW models,''
  arXiv:1311.4696 [hep-th].

\bibitem{BR}
   B.~Basso and A.~Rej,
  ``On the integrability of two-dimensional models with $U(1) \times SU(N)$ symmetry,''  
   Nucl.\ Phys.\ B {\bf 866} (2013) 337  [arXiv:1207.0413 [hep-th]]. 
 
\bibitem{DMV}
  F.~Delduc, M.~Magro and B.~Vicedo,
  ``On classical q-deformations of integrable sigma-models,''  
  JHEP {\bf 1311} (2013) 192  [arXiv:1308.3581 [hep-th]]. 
 
\bibitem{DMV2}
  F.~Delduc, M.~Magro and B.~Vicedo,
  ``An integrable deformation of the AdS$_5\times$S$^5$ superstring action,''  
 Phys.\ Rev.\ Lett.\  {\bf 112} (2014) 051601
  [arXiv:1309.5850 [hep-th]].
  
\bibitem{ABF}
    G.~Arutyunov, R.~Borsato and S.~Frolov,
  ``S-matrix for strings on $\eta$-deformed AdS$_5\times$S$^5$,''
  arXiv:1312.3542 [hep-th]. 

\bibitem{HRT}
  B.~Hoare, R.~Roiban and A.~A.~Tseytlin,
  ``On deformations of AdS$_n \times$S$^n$ supercosets,''
  arXiv:1403.5517 [hep-th].  

\bibitem{AdLvT}
G.~Arutynov, M.~de Leeuw and S.~J.~van Tongeren,
  ``On the exact spectrum and mirror duality of the $(AdS_5 \times S^5)_\eta$ superstring,''
  arXiv:1403.6104 [hep-th].  

%%%%%%%%%%%%%%%%%%%%%%
%   deformed S-matrix
%%%%%%%%%%%%%%%%%%%%%%  


\bibitem{BK}
  N.~Beisert and P.~Koroteev,
  ``Quantum Deformations of the One-Dimensional Hubbard Model,''  
J.\ Phys.\ A {\bf 41} (2008) 255204  [arXiv:0802.0777 [hep-th]]. \\ 
 N.~Beisert, W.~Galleas and T.~Matsumoto, 
``A Quantum Affine Algebra for the Deformed Hubbard Chain,'' 
J.\ Phys.\ A {\bf 45} (2012) 365206 [arXiv:1102.5700 [math-ph]].  

\bibitem{HHM}
 B.~Hoare, T.~J.~Hollowood and J.~L.~Miramontes,
  ``$q$-Deformation of the AdS$_5\times$S$^5$ Superstring S-matrix 
and its Relativistic Limit,''  
JHEP {\bf 1203} (2012) 015  [arXiv:1112.4485 [hep-th]]; 
  ``Bound States of the $q$-Deformed AdS$_5\times$S$^5$ Superstring S-matrix,''  
JHEP {\bf 1210} (2012) 076  [arXiv:1206.0010 [hep-th]]; 
 ``Restoring Unitarity in the $q$-Deformed World-Sheet S-Matrix,'' 
 JHEP {\bf 1310} (2013) 050 [arXiv:1303.1447 [hep-th]].

\bibitem{dLRT}
  M.~de Leeuw, V.~Regelskis and A.~Torrielli,
  ``The Quantum Affine Origin of the AdS/CFT Secret Symmetry,''  
J.\ Phys.\ A {\bf 45} (2012) 175202  [arXiv:1112.4989 [hep-th]].

\bibitem{Arutyunov}
  G.~Arutyunov, M.~de Leeuw and S.~J.~van Tongeren,
  ``The Quantum Deformed Mirror TBA I,''  
JHEP {\bf 1210} (2012) 090 [arXiv:1208.3478 [hep-th]];  
 ``The Quantum Deformed Mirror TBA II,''  
JHEP {\bf 1302} (2013) 012 [arXiv:1210.8185 [hep-th]]. 
   
%%% Jordanian of string action %%% 
  
\bibitem{KMY-Jordanian-typeIIB}
  I.~Kawaguchi, T.~Matsumoto and K.~Yoshida,
  ``Jordanian deformations of the AdS$_5\times$S$^5$ superstring,''
  arXiv:1401.4855 [hep-th]. 
  
\bibitem{SUGRA-KMY}
 I.~Kawaguchi, T.~Matsumoto and K.~Yoshida,
  ``A Jordanian deformation of AdS space in type IIB supergravity,''
  arXiv:1402.6147 [hep-th].  

\bibitem{KY-Sch}
  I.~Kawaguchi and K.~Yoshida,
  ``Classical integrability of Schr\"odinger sigma models and $q$-deformed Poincare symmetry,''  
JHEP {\bf 1111} (2011) 094  [arXiv:1109.0872 [hep-th]]; 
  ``Exotic symmetry and monodromy equivalence in Schr\"odinger sigma models,''  
JHEP {\bf 1302} (2013) 024  [arXiv:1209.4147 [hep-th]].

\bibitem{Jordanian-KMY}
  I.~Kawaguchi, T.~Matsumoto and K.~Yoshida,
  ``Schr\"odinger sigma models and Jordanian twists,''  
JHEP {\bf 1308} (2013) 013  [arXiv:1305.6556 [hep-th]].  

\bibitem{Kame}
  T.~Kameyama and K.~Yoshida,
  ``String theories on warped AdS backgrounds and integrable deformations of spin chains,''  
  JHEP {\bf 1305} (2013) 146  [arXiv:1304.1286 [hep-th]]. 

\bibitem{GS}
  D.~Giataganas and K.~Sfetsos,
  ``Non-integrability in non-relativistic theories,'' \\ 
  arXiv:1403.2703 [hep-th].

%%%%%%%%%%%%%%%%%%%%
  
\bibitem{R}
  N.~Reshetikhin,
  ``Multiparameter quantum groups and twisted quasitriangular Hopf algebras,''  
  Lett.\ Math.\ Phys.\  {\bf 20} (1990) 331.  
  

  
%%%%%%%% LS deformations %%%%%%%%  

\bibitem{LS}
  R.~G.~Leigh and M.~J.~Strassler,
  ``Exactly marginal operators and duality in four-dimensional N=1 supersymmetric gauge theory,''
  Nucl.\ Phys.\  B {\bf 447} (1995) 95
  [arXiv:hep-th/9503121].


\bibitem{LM}
 O.~Lunin and J.~M.~Maldacena,
  ``Deforming field theories with $U(1) \times U(1)$ global symmetry and their gravity duals,''  
JHEP {\bf 0505} (2005) 033  [hep-th/0502086].

\bibitem{Frolov}
  S.~Frolov,
  ``Lax pair for strings in Lunin-Maldacena background,''
  JHEP {\bf 0505} (2005) 069
  [hep-th/0503201].
  
\bibitem{B-Cherkis}
  D.~Berenstein and S.~A.~Cherkis,
  ``Deformations of N=4 SYM and integrable spin chain models,''  
Nucl.\ Phys.\ B {\bf 702} (2004) 49  [hep-th/0405215].  

\bibitem{HU}
J.~Fokken, C.~Sieg and M.~Wilhelm,
  ``Non-conformality of $\gamma_i$-deformed $\mathcal{N}=4$ SYM theory,''
  arXiv:1308.4420 [hep-th]; 
  ``The complete one-loop dilatation operator of planar real beta-deformed $\mathcal{N}=4$ SYM theory,''
  arXiv:1312.2959 [hep-th].

%%%%%%%%%%%%%%%%%%%%%%%%%%%%%%%%%%%%%%%%%%%%%%%%%%

\bibitem{AAF}
  L.~F.~Alday, G.~Arutyunov and S.~Frolov,
  ``Green-Schwarz strings in TsT-transformed backgrounds,''
  JHEP {\bf 0606} (2006) 018
  [hep-th/0512253].
  
\bibitem{DG}
  D.~Giataganas, L.~A.~Pando Zayas and K.~Zoubos,
  ``On Marginal Deformations and Non-Integrability,''
  JHEP {\bf 1401} (2014) 129
  [arXiv:1311.3241 [hep-th]].  
  
%%%%%%%%%%%%%%%%%%%%%%%%%%%%%%%%%%%%%%%%%%%%%%%%%%


  
\bibitem{Beauty}
  N.~Beisert and R.~Roiban,
  ``Beauty and the twist: The Bethe ansatz for twisted N=4 SYM,''
  JHEP {\bf 0508} (2005) 039
  [hep-th/0505187].  

\bibitem{ABBN}
  C.~Ahn, Z.~Bajnok, D.~Bombardelli and R.~I.~Nepomechie,
  ``Twisted Bethe equations from a twisted S-matrix,''
  JHEP {\bf 1102} (2011) 027
  [arXiv:1010.3229 [hep-th]]. 

\bibitem{AKL}
  C.~Ahn, M.~Kim and B.~-H.~Lee,
  ``Worldsheet S-matrix of beta-deformed SYM,''
  Phys.\ Lett.\ B {\bf 719} (2013) 458
  [arXiv:1211.4506 [hep-th]].  

%%%%%%%%%%%%%%%%%%%%%%%%%%%%%%%%%%%%%%%%%%%%%%%%%%%%
    
\bibitem{HRR}
  V.~E.~Hubeny, M.~Rangamani and S.~F.~Ross,
  ``Causal structures and holography,''
  JHEP {\bf 0507} (2005) 037
  [hep-th/0504034].  
    
\bibitem{future}
in preparation. 

\bibitem{AF-review}
  G.~Arutyunov and S.~Frolov,
  ``Foundations of the AdS$_5\times$S$^5$ Superstring. Part I,''
  J.\ Phys.\ A {\bf 42} (2009) 254003
  [arXiv:0901.4937 [hep-th]].      

\end{thebibliography}
\end{document}